# Complete Identification of a Dynamic Fractional Order System Under Non-ideal Conditions Using Fractional Differintegral Definitions

Deepyaman Maiti [#1], Ayan Acharya [#2], R. Janarthanan [*3], Amit Konar [#4]

[#] *Department of Electronics and Telecommunication Engineering, Jadavpur University*
*188 Raja S. C. Mallik Road, Kolkata - 700032, India*
[1] `deepyamanmaiti@gmail.com`
[2] `masterayan@gmail.com`
[4] `konaramit@yahoo.co.in`

[*] *Department of IT, Jaya Engineering College*
*C.T.H.Road, Thiruninravur, Chennai – 600024, India*
[3] `srmjana_73@yahoo.com`

*Abstract*—This contribution deals with identification of fractional-order dynamical systems. System identification, which refers to estimation of process parameters, is a necessity in control theory. Real processes are usually of fractional order as opposed to the ideal integral order models. A simple and elegant scheme of estimating the parameters for such a fractional order process is proposed. This method employs fractional calculus theory to find equations relating the parameters that are to be estimated, and then estimates the process parameters after solving the simultaneous equations. The data used for the calculations are intentionally corrupted to simulate real-life conditions. Results show that the proposed scheme offers a very high degree of accuracy even for erroneous data.

I. INTRODUCTION

Proper estimation of the parameters of a real process, fractional or otherwise, is a challenge to be encountered in the context of system identification [1], [2]. Accurate knowledge of the parameters of a system is often the first step in designing controllers. Many statistical and geometric methods such as least square and regression models are widely used for real-time parameter estimation.

The problem of parameter estimation becomes more difficult for a fractional order system compared to an integral order one. The real world objects or processes that we need to estimate are generally of fractional order [3]. A typical example of a non-integer (fractional) order system is the voltage-current relation of a semi-infinite lossy RC line or diffusion of heat into a semi-infinite solid, where the heat flow is equal to the half-derivative of temperature.

The fractional order of the system was earlier ignored because of the non-existence of simple mathematical tools for the description of such systems. Since major advances have been made in this area recently, it is possible to consider also the real order of the dynamical systems. Such models are more adequate for the description of dynamical systems with distributed parameters than integer-order models with concentrated parameters. With regard to this, in the task of identification, it is necessary to consider also the fractional-order of the dynamical system. Most classical identification methods cannot cope with fractional order transfer functions. Yet, this challenge must be overcome if we want to design a proper adaptive or self-tuning fractional order controller. As proved elsewhere [4] – [6], fractional order controllers are much superior to their traditional integral counterparts. Need for design of adaptive controllers gives an impetus to finding accurate schemes for system identification. This is the motivation for the present work.

Computation of transfer characteristics of the fractional order dynamic systems has been the subject of several publications: e.g. by numerical methods [7], as well as by analytical methods [8], using multi-layered neural networks [9], using the concept of continuous order-distribution [10], in the time domain using a state-space representation in [11], using the recursive algorithm approach in [12]. The application of fractional system identification to lead acid battery state of charge estimation was studied in [13]. In [14], a new modelling of electrical networks based on fractional order systems was proposed.

In this paper we propose a method for parameter identification of a fractional order system for a chosen structure of the model using fractional calculus theory to obtain simultaneous equations relating the unknown parameters and then solving these equations to obtain accurate estimates. This method enables us to work with the actual fractional order process rather than an integer order approximation. Using it in a system with known parameters will do the verification of the correctness of the identification.

We first consider that the fractional powers are constant and display the accuracy of the proposed method both when random corruptions are absent and present. Then we remove this limitation and propose two alternative schemes by which a fractional order system can be completely identified with a high degree of accuracy even in presence of significant quantities of error in the readings. It is necessary to understand the theory of fractional calculus in order to realize the significance of fractional order integration and derivation.

## II. FRACTIONAL CALCULUS THEORY

The fractional calculus is a generalization of integration and derivation to non-integer order operators. At first, we generalize the differential and integral operators into one fundamental operator $_aD_t^\alpha$ where:

$$_aD_t^\alpha = \frac{d^\alpha}{dt^\alpha} \text{ for } \Re(\alpha) > 0;$$
$$= 1 \text{ for } \Re(\alpha) = 0;$$
$$= \int_a^t (d\tau)^{-\alpha} \text{ for } \Re(\alpha) > 0. \quad (1)$$

$\Re(\alpha)$ being the real part of the order of differintegration.

The two definitions used for fractional differintegral are the Riemann-Liouville definition and the Grunwald-Letnikov definition [15].

The Riemann-Liouville definition is:

$$_aD_t^\alpha f(t) = \frac{1}{\Gamma(n-\alpha)} \frac{d^n}{dt^n} \int_a^t \frac{f(\tau)}{(t-\tau)^{\alpha-n+1}} d\tau \quad (2)$$

for $(n-1 < \alpha < n)$ and $\Gamma(x)$ is Euler's gamma function.

The Laplace transform method is used for solving engineering problems. The formula for the Laplace transform of the Riemann-Liouville fractional derivative (2) has the form:

$$\int_0^\infty e^{-pt} {_0D_t^\alpha} f(t) dt = p^\alpha F(p) - \sum_{k=0}^{n-1} p^k {_0D_t^{\alpha-k-1}} f(t)\big|_{t=0}$$

for $(n-1 < \alpha \leq n)$.

The Grunwald-Letnikov definition is:

$$_aD_t^\alpha f(t) = \lim_{h \to 0} \frac{1}{h^\alpha} \sum_{j=0}^{\left[\frac{t-a}{h}\right]} (-1)^j \binom{\alpha}{j} f(t-jh) \quad (3)$$

where $[y]$ means the integer part of y.

Derived from the Grunwald-Letnikov definition, the numerical calculation formula of fractional derivative can be achieved as $_{t-L}D_t^\alpha x(t) \approx h^{-\alpha} \sum_{j=0}^{[L/T]} b_j x(t-jh) \quad (4)$ where L is the length of memory. T, the sampling time always replaces the time increment h during approximation. The weighting coefficients $b_j$ can be calculated recursively by

$$b_0 = 1, b_j = \left(1 - \frac{1+\alpha}{j}\right) b_{j-1}, (j \geq 1) \quad (5)$$

## III. PROCESS OF IDENTIFICATION WHEN FRACTIONAL POWERS ARE CONSTANT

We have considered a five-parameter model of a fractional order process whose transfer function is of the form $\frac{1}{a_1 s^\alpha + a_2 s^\beta + a_3}$. The orders of fractionality $\alpha$ and $\beta$ are known and the coefficients $a_1$, $a_2$ and $a_3$ are to be estimated. One important advantage of the proposed scheme is that we do not require to know the ranges of variation of $a_1$, $a_2$ and $a_3$. It should be noted that without loss of generality, we may presume the dc gain to be unity so that the dc gain and its possible fluctuations are included in the coefficients $a_1$, $a_2$ and $a_3$. If C(s) is the output and R(s) the input,

$$\frac{C(s)}{R(s)} = \frac{1}{a_1 s^\alpha + a_2 s^\beta + a_3},$$
$$\Rightarrow R(s) = a_1 s^\alpha C(s) + a_2 s^\beta C(s) + a_3 C(s).$$

In time domain,

$$r(t) = a_1 D^\alpha c(t) + a_2 D^\beta c(t) + a_3 c(t) \quad (6)$$

$$\Rightarrow r(t) \approx a_1 T^{-\alpha} \sum_{j=0}^{[L/T]} b_j c(t-jT) + a_2 T^{-\beta} \sum_{j=0}^{[L/T]} b_j c(t-jT) + a_3 c(t) \quad (7)$$

The proposed scheme requires sampled input at time instant t and sampled outputs at time instants t, t – T, t – 2T, t – 3T, .......... Sampled outputs are required for a time length L previous to t, T being the sampling time. Calculation of fractional derivatives and integrals requires the past history of the process to be remembered. So more the value of L, the better.

Thus the values of $D^\alpha c(t)$ and $D^\beta c(t)$ can be calculated so that (6) reduces to the form $a_1 p + a_2 q + a_3 r = s$, where $p, q, r, s$ are constants whose values have been determined.

Let us assume that we have a set of sampled outputs c(t) from the system for unit step test signal.

That is, we have

$$u(t) = a_1 D^\alpha c(t) + a_2 D^\beta c(t) + a_3 c(t). \quad (8)$$

Now there are three unknown parameters, namely $a_1$, $a_2$ and $a_3$. So we need three simultaneous equations to solve from them. One equation is (8). We will integrate both sides of (8) to obtain $\int u(t) dt = \int [a_1 D^\alpha c(t) + a_2 D^\beta c(t) + a_3 c(t)] dt$

which gives us

$$r(t) = a_1 D^{\alpha-1} c(t) + a_2 D^{\beta-1} c(t) + a_3 D^{-1} c(t) \quad (9)$$

where r(t) signifies unit ramp input and c(t) is the output due to unit step input. Thus we have derived a second equation relating $a_1$, $a_2$ and $a_3$.

The third equation will be obtained by integrating both sides of (9). This gives us

$$p(t) = a_1 D^{\alpha-2} c(t) + a_2 D^{\beta-2} c(t) + a_3 D^{-2} c(t) \quad (10)$$

where p(t) signifies parabolic input and c(t) is the output due to unit step input.

It can be seen that (8), (9), (10) are distinct equations in $a_1$, $a_2$ and $a_3$. So we can solve them simultaneously to identify the three unknown parameters $a_1$, $a_2$ and $a_3$.

As we have displayed elsewhere, direct application of the above scheme gives very satisfactory results when the readings c(t) are accurate. If we now add an error component e(t) to c(t) to have a distorted output waveform $c(t) \equiv c(t) + e(t)$ from which we want to make our identification, (8) will be transformed to
$$u(t) = a_1 D^{\alpha}[c(t)+e(t)] + a_2 D^{\beta}[c(t)+e(t)] + a_3[c(t)+e(t)] \quad (11)$$

So (11) will not give an accurate relation between $a_1$, $a_2$ and $a_3$ due to the presence of the terms $a_1 D^{\alpha} e(t)$, $a_2 D^{\beta} e(t)$ and $a_3 e(t)$. Hence, the equations obtained by applying the transformation $c(t) \equiv c(t) + e(t)$ on (9) and (10) will also be inaccurate. Our aim will be to minimize this inaccuracy.

One significant fact we observed is that for the same random error waveform e(t), $D^{\alpha_1} e(t) \ll D^{\alpha_2} e(t)$ if $\alpha_1 < 0$ and $\alpha_2 > 0$ when, in effect, $D^{\alpha_1} e(t)$ becomes an integration.

Although we are not yet ready to put forward a rigorous mathematical proof for this observation, from (4) we see that $D^{\alpha} e(t)$ contains a factor $h^{-\alpha}$, where the sampling interval h << 1. Hence $h^{-\alpha} \gg 1$ for $\alpha > 0$ and $h^{-\alpha} \ll 1$ for $\alpha < 0$. Of course $b_j$ contains $\alpha$ and this will have an impact on the value of $D^{\alpha} e(t)$. But it can be easily seen that the effect due to $h^{-\alpha}$ is much more significant.

To support our contention, in Table I, we tabulate the values of $D^{\alpha} e(t)$ for 10 different sets of e(t) with $\alpha$ = 1.5, 1.2, 0.9, 0.6, 0.3, -0.3, -0.6, -0.9, -1.2, -1.5. The amplitude of e(t) varies between –0.01 and 0.01. Length of memory = 10 seconds, i.e. the fractional derivatives are calculated at time t = 10 seconds. Sampling rate is once in 0.001 seconds.

The transfer function of our system is $\frac{1}{a_1 s^{\alpha} + a_2 s^{\beta} + a_3}$, and as we are well aware, $\alpha, \beta > 0$ for a practical system, so that in (11), the orders of derivation $\alpha, \beta$ are positive.

To remedy this, we can perform a simple transformation on the transfer function of the system, which we can write as
$\frac{s^{-n}}{a_1 s^{\alpha-n} + a_2 s^{\beta-n} + a_3 s^{-n}}$, where $(n-1) < \alpha < n$ and $\alpha > \beta$.

Proceeding as before we can now obtain our three simultaneous equations as:
$$D^{-n} u(t) = (a_1 D^{\alpha-n} + a_2 D^{\beta-n} + a_3 D^{-n})[c(t)+e(t)] \quad (12)$$
$$D^{-n-1} u(t) = (a_1 D^{\alpha-n-1} + a_2 D^{\beta-n-1} + a_3 D^{-n-1})[c(t)+e(t)] \quad (13)$$
$$D^{-n-2} u(t) = (a_1 D^{\alpha-n-2} + a_2 D^{\beta-n-2} + a_3 D^{-n-2})[c(t)+e(t)] \quad (14)$$

It can now be checked that all orders of derivation are now negative so that we will actually be performing fractional order integrations rather than fractional order differentiations.

IV. ILLUSTRATION

Let the process whose parameters are to be estimated be $\frac{1}{a_1 s^{2.23} + a_2 s^{0.88} + a_3}$.

The input considered is r(t) = 1 i.e. unit step.

Synthetic data for c(t) is created using $a_1 = 0.8$, $a_2 = 0.5$ and $a_3 = 1$, i.e. the values of c(t) are obtained at different time instants (using a MATLAB program) assuming a process with transfer function $\frac{1}{0.8 s^{2.23} + 0.5 s^{0.88} + 1}$. The simultaneous equations corresponding to (12), (13) and (14) are
$$D^{-3} u(t) = (a_1 D^{-0.77} + a_2 D^{-2.12} + a_3 D^{-3})[c(t)+e(t)] \quad (15)$$
$$D^{-4} u(t) = (a_1 D^{-1.77} + a_2 D^{-3.12} + a_3 D^{-4})[c(t)+e(t)] \quad (16)$$
$$D^{-5} u(t) = (a_1 D^{-2.77} + a_2 D^{-4.12} + a_3 D^{-5})[c(t)+e(t)] \quad (17)$$

Length of memory L = 10 seconds and T = 0.001 seconds is used to calculate the fractional derivatives.

We will display the accuracy of identification when the output readings used to calculate the fractional derivatives are ideal and also when they are erroneous to the extent of a random error component in the range [−0.05,0.05] in each reading. This error component is quite large since the output response is often below unity. The output response of the system for unit step input is given is Fig. 1 both in presence and absence of the error component.

TABLE I.

VARIATION OF $D^{\alpha}$e(t) WITH α. (THE 10 SEQUENCES e(t) ARE CONSECUTIVE AND INDEPENDENT.)

| e(t) | $D^{\alpha}$ e(t) for derivation order $\alpha$ | | | | | | | | | |
|---|---|---|---|---|---|---|---|---|---|---|
| | α = 1.5 | α = 1.2 | α = 0.9 | α = 0.6 | α = 0.3 | α = -0.3 | α = -0.6 | α = -0.9 | α = -1.2 | α = -1.5 |
| 1 | -435.7842 | -50.7575 | -5.7583 | -0.6287 | -0.0661 | -0.0008 | 0.0001 | 0.0006 | 0.0013 | 0.0025 |
| 2 | -603.6659 | -59.5517 | -5.7933 | -0.5742 | -0.0617 | -0.0013 | -0.0005 | -0.0004 | -0.0002 | 0.0002 |
| 3 | 424.4136 | 44.8209 | 4.7948 | 0.5242 | 0.0581 | 0.0002 | -0.0002 | -0.0001 | -0.0001 | -0.0003 |
| 4 | -256.3730 | -26.5634 | -3.0495 | -0.3928 | -0.0549 | -0.0011 | -0.0002 | -0.0001 | -0.0002 | -0.0005 |
| 5 | -107.8138 | -12.0636 | -1.4119 | -0.1631 | -0.0164 | 0.0004 | 0.0004 | 0.0005 | 0.0008 | 0.0013 |
| 6 | 642.4164 | 78.0679 | 9.1798 | 1.0311 | 0.1069 | 0.0006 | -0.0001 | -0.0002 | -0.0006 | -0.0012 |
| 7 | 184.7026 | 22.6486 | 2.6896 | 0.3051 | 0.0321 | 0.0000 | -0.0001 | -0.0001 | 0.0002 | 0.0003 |
| 8 | -393.9215 | -45.1151 | -5.0296 | -0.5467 | -0.0562 | 0.0001 | 0.0002 | 0.0000 | -0.0001 | -0.0002 |
| 9 | -109.5421 | -4.2756 | 0.4383 | 0.1517 | 0.0272 | 0.0006 | 0.0001 | 0.0000 | -0.0001 | 0.0000 |
| 10 | -32.4628 | -7.6152 | -1.4990 | -0.2670 | -0.0439 | -0.0008 | 0.0002 | 0.0007 | 0.0014 | 0.0024 |

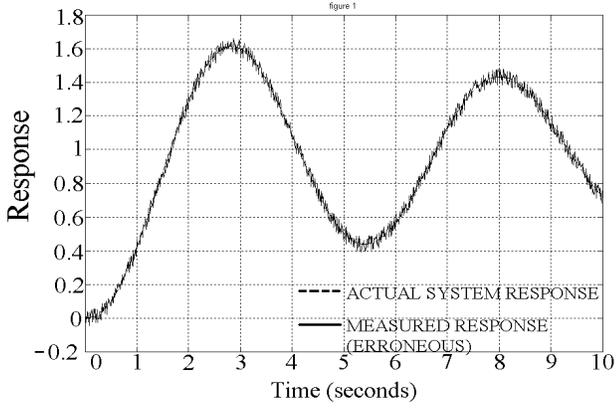

Fig. 1. Unit step response of the system in presence and absence of error component

### A. Ideal Case: e(t) = 0 for all t

The following derivatives are then calculated numerically using (4) and (5):

$D^{-0.77}c(t) = 6.1777$,     $D^{-2.12}c(t) = 51.3011$,

$D^{-3}c(t) = 136.1477$,     $D^{-1.77}c(t) = 32.2818$,

$D^{-3.12}c(t) = 152.6826$,     $D^{-4}c(t) = 314.8183$,

$D^{-2.77}c(t) = 108.0207$,     $D^{-4.12}c(t) = 342.4005$,

$D^{-5}c(t) = 576.6986$.

The set of simultaneous equations is

$$\begin{bmatrix} 6.1777 & 51.3011 & 136.1477 \\ 32.2818 & 152.6826 & 314.8183 \\ 108.0207 & 342.4005 & 576.6986 \end{bmatrix} \begin{bmatrix} a_1 \\ a_2 \\ a_3 \end{bmatrix} = \begin{bmatrix} 166.7167 \\ 416.9167 \\ 834.1670 \end{bmatrix}$$

After solving we have $a_1 = 0.8001, a_2 = 0.4996, a_3 = 1.0000$ as the unknown parameters. The errors in estimating them are respectively 0.0125%, 0.0800% and 0%. Summation of square errors of this process model outputs relative to the output data set for unit step input is 0.0030.

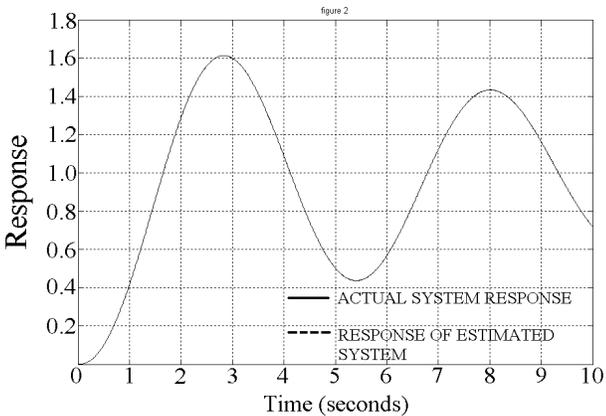

Fig. 2. Unit step responses of actual and estimated system for sub-section A (ideal case)

### B. Non-ideal Case: Each Element in e(t) is Between –0.05 and 0.05

To each reading c(t) is added a random error component varying between –0.05 and 0.05.

We proceed as before to obtain the set of simultaneous equations as

$$\begin{bmatrix} 6.1798 & 51.3179 & 136.1948 \\ 32.2919 & 152.7357 & 314.9314 \\ 108.0577 & 342.5242 & 576.9207 \end{bmatrix} \begin{bmatrix} a_1 \\ a_2 \\ a_3 \end{bmatrix} = \begin{bmatrix} 166.7167 \\ 416.9167 \\ 834.1670 \end{bmatrix}$$

After solving we have $a_1 = 0.7992, a_2 = 0.4996, a_3 = 0.9996$ as the unknown parameters. The errors in estimating them are respectively 0.1000%, 0.0800% and 0.0400%.

Summation of square errors of this process model outputs relative to the actual output data set for unit step input is 0.0062.

The unit step responses of the actual and the estimated systems are shown in Fig. 3 for this particular case.

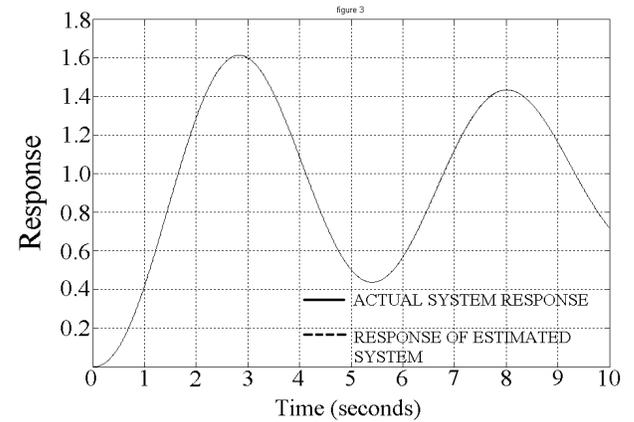

Fig. 3. Unit step responses of actual and estimated system for sub-section B (non-ideal case)

## V. PROCESS OF IDENTIFICATION WHEN FRACTIONAL POWERS ARE VARYING

We shall propose two algorithms for system identification for this situation and then perform a comparative analysis between them. Since we are satisfied that we can accurately identify the coefficient terms $a_1$, $a_2$ and $a_3$ when the fractional powers α and β are known, these two alternative algorithms will concentrate on first searching the fractional power state-space. Once α and β has been identified, the coefficients will be calculated. We will generate possible process models according to some criteria and then choose the optimum model(s) based on their time responses as obtained by simulations.

Let us first define a fitness parameter $F = \sum_{t=i}^{f}[f(t)-g(t)]^2$ for each process model where f(t) is the set of outputs obtained from the actual process and g(t) is the set of outputs generated by a possible process model at the corresponding time instants. The lower the value of F for a model, the better it is.

*A. Algorithm 1*

We will consider that the fractional power α can vary in the range $\{\alpha_{min}, \alpha_{max}\}$ and β can vary in the range $\{\beta_{min}, \beta_{max}\}$. We will subdivide each of the two ranges $(\alpha_{max}-\alpha_{min})$ and $(\beta_{max}-\beta_{min})$ into m and n intervals respectively. (m may or may not be equal to n, depending on our design of the interval lengths.) We will consider as the nominal value of α and β for each interval its central point.

That is, we may have m possible values of α:
$\alpha_{min}+\frac{(\alpha_{max}-\alpha_{min})}{2m}$, $\alpha_{min}+\frac{3(\alpha_{max}-\alpha_{min})}{2m}$,
$\alpha_{min}+\frac{5(\alpha_{max}-\alpha_{min})}{2m}$, $\alpha_{min}+\frac{7(\alpha_{max}-\alpha_{min})}{2m}$,
........., $\alpha_{min}+\frac{(2m-1)(\alpha_{max}-\alpha_{min})}{2m}$.

Likewise we will have n possible values of β:
$\beta_{min}+\frac{(\beta_{max}-\beta_{min})}{2n}$, $\beta_{min}+\frac{3(\beta_{max}-\beta_{min})}{2n}$,
$\beta_{min}+\frac{5(\beta_{max}-\beta_{min})}{2n}$, $\beta_{min}+\frac{7(\beta_{max}-\beta_{min})}{2n}$,
........., $\beta_{min}+\frac{(2n-1)(\beta_{max}-\beta_{min})}{2n}$.

Thus the fractional powers $(\alpha, \beta)$ may assume any of the m x n values (or be sufficiently close to be an acceptable estimate). Now, for each of the m x n possible fractional power combinations we will find out the coefficients $a_1$, $a_2$ and $a_3$ by the method of forming and solving simultaneous equations as illustrated in sections III and IV.

Thus we will have m x n possible process models. For each of these we will calculate the fitness F. We will select the process model with the least value of F, i.e. with highest fitness.

*B. Illustration of Algorithm 1 under Non-ideal case: Each Element in e(t) is Between –0.05 and 0.05*

Assume that α varies from 2.0 to 2.4, i.e. $\alpha_{min}=2.0$ and $\alpha_{max}=2.4$; β varies from 0.7 to 1.1, i.e. $\beta_{min}=0.7$ and $\beta_{max}=1.1$. Also we let m = n = 20.

Therefore the 20 possible values of α are 2.01, 2.03, 2.05, ......., 2.37 and 2.39. The 20 possible values for β are 0.71, 0.73, 0.75, ......., 1.07 and 1.09.

Then we compute the solution sets of $a_1, a_2, a_3$ for each of the 400 $(\alpha, \beta)$ values and also note the fitness F of each process model. The process model with the best F is our estimated system model. The estimated parameters and the fitness values are tabulated in Table II for the 10 best models.

TABLE II
BEST 10 IDENTIFICATIONS AS PER ALGORITHM 1

| Sl. No. | Estimated Parameters | | | | | Fitness F |
|---|---|---|---|---|---|---|
| | α | β | $a_1$ | $a_2$ | $a_3$ | |
| 1 | 2.23 | 0.87 | 0.8039 | 0.4979 | 0.9975 | 0.4036 |
| 2 | 2.23 | 0.89 | 0.7933 | 0.5017 | 1.0013 | 0.4901 |
| 3 | 2.21 | 0.85 | 0.8141 | 0.4866 | 0.9952 | 1.0209 |
| 4 | 2.25 | 0.91 | 0.7823 | 0.5135 | 1.0036 | 1.4563 |
| 5 | 2.21 | 0.83 | 0.8238 | 0.4836 | 0.9911 | 1.7270 |
| 6 | 2.25 | 0.93 | 0.7709 | 0.5183 | 1.0071 | 2.8737 |
| 7 | 2.21 | 0.87 | 0.8040 | 0.4900 | 0.9990 | 3.2772 |
| 8 | 2.23 | 0.85 | 0.8141 | 0.4944 | 0.9936 | 3.8235 |
| 9 | 2.23 | 0.91 | 0.7824 | 0.5060 | 1.0049 | 3.9228 |
| 10 | 2.25 | 0.89 | 0.7934 | 0.5092 | 0.9999 | 3.9810 |

As we can see, the estimated model is the model with the best fitness, i.e. the first model. The percentage errors in identification of α, β, $a_1$, $a_2$, $a_3$ are respectively 0, 1.1364, 0.4875, 0.4200, 0.2500.

The unit step responses for the actual and the best-estimated system are given below in Fig. 4.

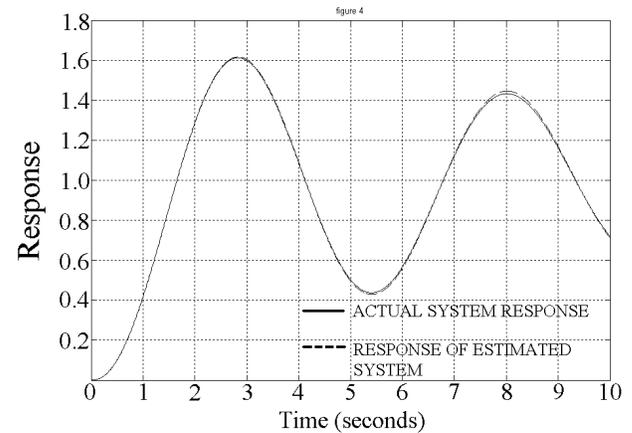

Fig. 4. Unit step responses of actual and best-estimated system for algorithm 1

*C. Algorithm 2*

As in algorithm 1, we will consider that the fractional power α can vary in the range $\{\alpha_{min}, \alpha_{max}\}$ and β can vary in the range $\{\beta_{min}, \beta_{max}\}$. We will subdivide each of the two ranges $(\alpha_{max}-\alpha_{min})$ and $(\beta_{max}-\beta_{min})$ into $m_1$ and $n_1$ intervals respectively. ($m_1$ may or may not be equal to $n_1$, depending on our design of the interval lengths.) We will consider as the nominal value of α and β for each interval as its central point.

Thus the fractional powers (α, β) may assume any of the $m_1$ x $n_1$ values (or be sufficiently close to be an acceptable estimate). Now for each possible fractional power combinations we will find out the coefficients $a_1$, $a_2$ and $a_3$ by the method of forming and solving simultaneous equations as illustrated in sections III and IV. Thus we will have $m_1$ x $n_1$ possible process models. For each of these we will calculate the fitness F through simulation. This completes one sub-run.

Let us define the concept of a temporary memory space (buffer) where we will store the best $p_1$ fitness values and the corresponding models. (The choice of $p_1$ depends on us. Obviously $p_1 \geq 1$.)

Corresponding to the $p_1$ models, we will have $p_1$ sub-intervals where α, β may lie. We will sub-divide each of the α-intervals and β-intervals into a suitable number of sub-intervals, say $m_2$ and $n_2$. So now we have $p_1$ x $m_2$ x $n_2$ possible process models. Once again, we shall compute the values of F and store the best $p_2$ values in the buffer. This completes the second sub-run.

In this way, we shall continue the sub-runs until we are satisfied that the value of F is sufficiently good.

The advantage of this algorithm (algorithm 2) is two-fold.

Firstly, it allows us to search more thoroughly within an interval because the search is performed not once but many times. So, we do not have to waste resources searching in an interval that consistently gives poor fitness values.

Secondly, because the search is many-layered, for each sub-run we can take low values of m and n and yet obtain better fitness. So this algorithm is more efficient and is of an adaptive nature.

Discussions for further improvement of this algorithm will be put forward after the illustration.

*D. Illustration of Algorithm 2 under Non-ideal Case: Each Element in e(t) is Between –0.05 and 0.05*

Assume that α varies from 2.0 to 2.4, i.e. $\alpha_{min} = 2.0$ and $\alpha_{max} = 2.4$; β varies from 0.7 to 1.1, i.e. $\beta_{min} = 0.7$ and $\beta_{max} = 1.1$.

1) *Sub-run 1*: $m_1 = n_1 = 4$. The α- nominal values are 2.05, 2.15, 2.25 and 2.35. The β- nominal values are 0.75, 0.85, 0.95 and 1.05. So we have 16 sets of α, β. We compute the 16 corresponding fitnesses. The best 4 models are kept in the buffer, i.e. $p_1 = 4$. The 4 best models for the sub-run 1 are:

TABLE III
BEST 4 MODELS FOR SUB-RUN 1

| Sl. No. | Estimated Parameters | | | | | Fitness F |
|---|---|---|---|---|---|---|
| | α | β | $a_1$ | $a_2$ | $a_3$ | |
| 1 | 2.25 | 0.95 | 0.7591 | 0.5235 | 1.0104 | 8.0226 |
| 2 | 2.15 | 0.75 | 0.8598 | 0.4502 | 0.9792 | 17.8231 |
| 3 | 2.25 | 0.85 | 0.8145 | 0.5021 | 0.9921 | 21.6413 |
| 4 | 2.15 | 0.85 | 0.8162 | 0.4614 | 1.0001 | 50.1798 |

The unit step responses of the actual and the first model (best-estimated at this stage) are given in Fig. 5.

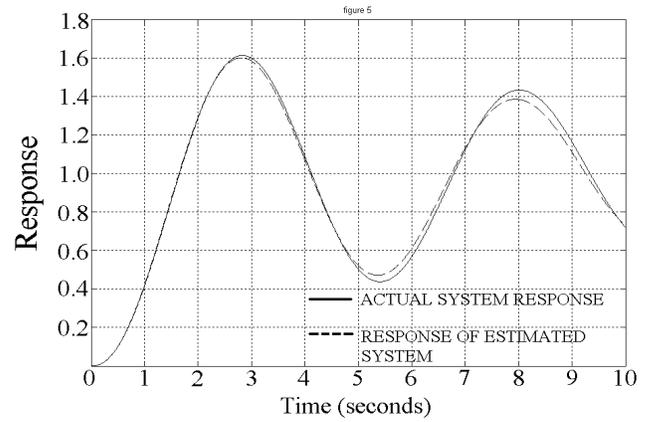

Fig. 5. Unit step responses of actual and the best-estimated system after sub-run 1, algorithm 2

The 4 α- and β- subintervals to search more thoroughly are: (2.2-2.3, 0.9-1.0), (2.1-2.2, 0.7-0.8), (2.2-2.3, 0.8-0.9) and (2.1-2.2, 0.8-0.9).

2) *Sub-run 2*: Each α- and β-sub-interval is divided into 5 sub-sub-intervals, i.e. $m_2 = n_2 = 5$. Hence there will be $p_1$ x $m_2$ x $n_2$ = 4 x 5 x 5 = 100 process models. The α- and β- nominal values are calculated as the central points of the respective intervals. For example, for the α- and β-sub-interval (2.2-2.3, 0.9-1.0), the α- nominal values are 2.21, 2.23, 2.25, 2.27 and 2.29. The β- nominal values are 0.91, 0.93, 0.95, 0.97 and 0.99. So we have 100 sets of α, β. We compute the 100 corresponding fitnesses. The best 3 models are kept in the buffer, i.e. $p_2 = 3$. The 3 best models are:

TABLE IV
BEST 3 MODELS FOR SUB-RUN 2

| Sl. No. | Estimated Parameters | | | | | Fitness F |
|---|---|---|---|---|---|---|
| | α | β | $a_1$ | $a_2$ | $a_3$ | |
| 1 | 2.23 | 0.87 | 0.8039 | 0.4979 | 0.9975 | 0.4036 |
| 2 | 2.23 | 0.89 | 0.7933 | 0.5017 | 1.0013 | 0.4901 |
| 3 | 2.21 | 0.85 | 0.8141 | 0.4866 | 0.9952 | 1.0209 |

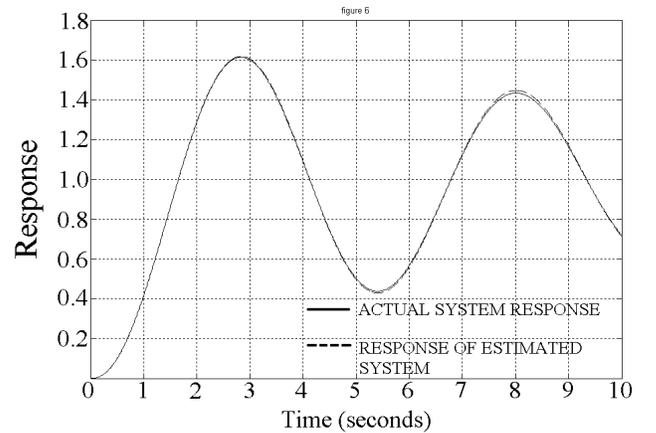

Fig. 6. Unit step responses of actual and the best-estimated system after sub-run 2, algorithm 2

The 3 α- and β- subintervals to search more thoroughly are: (2.22-2.24, 0.86-0.88), (2.22-2.24, 0.88-0.90) and (2.20-2.22, 0.84-0.86).

3) *Sub-run 3*: Each α- and β-sub-interval is divided into 5 sub-sub-intervals, i.e. $m_3 = n_3 = 5$. Hence there will be $p_2 \times m_3 \times n_3 = 3 \times 5 \times 5 = 75$ process models. The α- and β-nominal values are calculated as the central points of the respective intervals. So we have 75 sets of α, β. We compute the 75 corresponding fitnesses. The best 3 models are kept in the buffer, i.e. $p_3 = 3$. The 3 best models are:

TABLE V
BEST 3 MODELS FOR SUB-RUN 3

| Sl. No. | Estimated Parameters | | | | | Fitness F |
|---|---|---|---|---|---|---|
| | α | β | $a_1$ | $a_2$ | $a_3$ | |
| 1 | 2.230 | 0.878 | 0.8014 | 0.5000 | 0.9994 | 0.0205 |
| 2 | 2.234 | 0.886 | 0.7972 | 0.5031 | 1.0006 | 0.0228 |
| 3 | 2.230 | 0.882 | 0.7993 | 0.5008 | 1.0001 | 0.0252 |

If we decide to stop at this point, the estimated model is the model with the best fitness, i.e. the first model.

The percentage errors in identification of α, β, $a_1$, $a_2$, $a_3$ are respectively 0, 0.2273, 0.1750, 0, 0.0600.

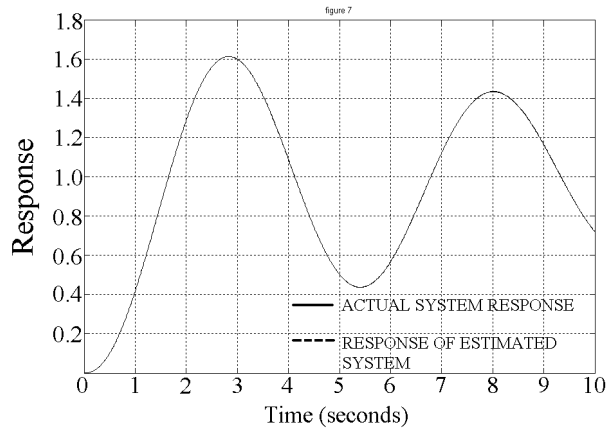

Fig. 7. Unit step responses of actual and the best-estimated system after sub-run 3, algorithm 2

### E. Relative Merits of Algorithm 2 over Algorithm 1

First and foremost, we can easily see from the percentage errors of estimated parameters that algorithm 2 yields much better results.

Secondly, for algorithm 1, we considered 20 x 20 = 400 process models. For algorithm 2, we needed to consider only 191 models. Thus algorithm 2 is much more efficient. This happens because in algorithm 2, we first take care to identify the possible sub-ranges of the fractional powers and then fine-search within probable ranges. Algorithm 1 employs only one level of searching, while algorithm 2 is adaptive and employs intensive searching within specific intervals.

Algorithm 1 is of course simpler to implement, but that remains its only advantage. Algorithm 2 offers better results and is more efficient.

### F. Possible Improvements in Algorithm 2

The way sub-runs 1, 2 and 3 in algorithm 2 were implemented was as follows. Suppose in a given sub-interval, α assumed the nominal values $\alpha_1, \alpha_2, \alpha_3, \alpha_4$ and $\alpha_5$. In the same sub-interval, β assumed the nominal values $\beta_1, \beta_2, \beta_3, \beta_4$ and $\beta_5$. We first fixed α at $\alpha_1$, varied β from $\beta_1$ to $\beta_5$ and noted down the fitness values. Then we changed α to $\alpha_2$, varied β from $\beta_1$ to $\beta_5$ and noted down the fitness values. In this way we continued until $\alpha = \alpha_5$.

A pattern was noticed in the corresponding fitness values. Either the fitness values decreased consistently from $\beta = \beta_1$ to $\beta = \beta_5$, or increased consistently, or first decreased then increased. That is to say, a definite pattern was followed. This pattern was noticed in *all* the searches. Obviously similar patterns in fitness values would have also been noted if we kept β fixed at $\beta_1$ and varied α from $\alpha_1$ to $\alpha_5$ etc.

TABLE VIA
FIRST PATTERN IN FITNESS VALUES

| α, β | $a_1$ | $a_2$ | $a_3$ | Fitness |
|---|---|---|---|---|
| 2.25, 0.81 | 0.8344 | 0.4967 | 0.9834 | 57.3075 |
| 2.25, 0.83 | 0.8246 | 0.4992 | 0.9878 | 37.1401 |
| 2.25, 0.85 | 0.8145 | 0.5021 | 0.9921 | 21.6413 |
| 2.25, 0.87 | 0.8041 | 0.5054 | 0.9961 | 10.6469 |
| 2.25, 0.89 | 0.7934 | 0.5092 | 0.9999 | 3.9810 |

TABLE VIB
SECOND PATTERN IN FITNESS VALUES

| α, β | $a_1$ | $a_2$ | $a_3$ | Fitness |
|---|---|---|---|---|
| 2.19, 0.81 | 0.8333 | 0.4728 | 0.9887 | 4.0644 |
| 2.19, 0.83 | 0.8240 | 0.4754 | 0.9928 | 5.5932 |
| 2.19, 0.85 | 0.8144 | 0.4785 | 0.9968 | 9.6201 |
| 2.19, 0.87 | 0.8045 | 0.4820 | 1.0006 | 16.0432 |
| 2.19, 0.89 | 0.7942 | 0.4859 | 1.0042 | 24.7542 |

TABLE VIC
THIRD PATTERN IN FITNESS VALUES

| α, β | $a_1$ | $a_2$ | $a_3$ | Fitness |
|---|---|---|---|---|
| 2.27, 0.91 | 0.7826 | 0.5208 | 1.0023 | 11.9583 |
| 2.27, 0.93 | 0.7710 | 0.5255 | 1.0059 | 6.3391 |
| 2.27, 0.95 | 0.7591 | 0.5306 | 1.0093 | 5.3533 |
| 2.27, 0.97 | 0.7467 | 0.5363 | 1.0125 | 8.7405 |
| 2.27, 0.99 | 0.7340 | 0.5424 | 1.0157 | 16.2263 |

Now, algorithm 2 can be made even more adaptive and efficient by simply checking whether the fitness value gets better or worse with changing β for fixed α. If it gets better, then we should continue changing β and noting the fitness values. If it is getting worse, we should move on to the next α value. In this fashion, the number of process models generated by algorithm 2 can be reduced by a factor close to two.

## VI. COMPARISON, COMMENTS AND CONCLUSIONS

An elegant method for the identification of parameters of a fractional order system is proposed. For very accurate results, value of L should be large and that of T small while calculating the fractional derivation numerically.

The challenge in fractional order system identification is that the fractional powers are not restricted to assume only discrete integral values, but are distributed in a continuous interval. For two integral order systems, identical time domain responses mean identical transfer functions. But for fractional order systems, we often find that a better identification of the actual process has actually a lower fitness than a worse model.

The method of finding a relation between the coefficients by use of fractional calculus renders the application of a complex evolutionary algorithm redundant. Therein lies its merit.

In the future, we plan to devise more efficient algorithms to scour the fractional powers state-space, so that higher order systems can be identified faster. We are also investigating the possibility of employing stochastic optimization algorithms for this purpose.